\documentclass[%
 reprint,
superscriptaddress,
 amsmath,amssymb,
 aps,
 prappl,
]{revtex4-2}

\usepackage{textcomp}
\usepackage{graphicx}
\usepackage{dcolumn}
\usepackage{bm}
\usepackage{upgreek}
\graphicspath{{images/}}

\begin{document}

\preprint{APS/123-QED}

\title{Symmetry of the Magnetoelastic Interaction of Rayleigh and Shear Horizontal Magnetoacoustic Waves in Nickel Thin Films on LiTaO$_3$}

\author{M. K{\"u}{\ss}}
 \email{matthias.kuess@physik.uni-augsburg.de.}
 \affiliation{Experimental Physics I, Institut of Physics, University of Augsburg, 86135 Augsburg, Germany\looseness=-1}
 
\author{M. Heigl}
 \affiliation{Experimental Physics IV, Institut of Physics, University of Augsburg, 86135 Augsburg, Germany\looseness=-1}

\author{L. Flacke}
 \affiliation{Walther-Mei{\ss}ner-Institut, Bayerische Akademie der Wissenschaften, 85748 Garching, Germany}
 \affiliation{Physics-Department, Technical University Munich, 85748 Garching, Germany}
 
 \author{A. Hefele}
 \affiliation{Experimental Physics I, Institut of Physics, University of Augsburg, 86135 Augsburg, Germany\looseness=-1}

\author{A. H{\"o}rner}
 \affiliation{Experimental Physics I, Institut of Physics, University of Augsburg, 86135 Augsburg, Germany\looseness=-1}

\author{M. Weiler}
 \affiliation{Walther-Mei{\ss}ner-Institut, Bayerische Akademie der Wissenschaften, 85748 Garching, Germany}
 \affiliation{Physics-Department, Technical University Munich, 85748 Garching, Germany}
 \affiliation{Fachbereich Physik and Landesforschungszentrum OPTIMAS, Technische Universit{\"a}t Kaiserslautern, 67663 Kaiserslautern, Germany}
 
\author{M. Albrecht}
 \affiliation{Experimental Physics IV, Institut of Physics, University of Augsburg, 86135 Augsburg, Germany\looseness=-1}
 
\author{A. Wixforth}
 \affiliation{Experimental Physics I, Institut of Physics, University of Augsburg, 86135 Augsburg, Germany\looseness=-1}


\date{\today}

\begin{abstract}

We study the interaction of Rayleigh and shear horizontal surface acoustic waves (SAWs) with spin waves in thin Ni films on a piezoelectric LiTaO$_3$ substrate, which supports both SAW modes simultaneously. Because Rayleigh and shear horizontal modes induce different strain components in the Ni thin films, the symmetries of the magnetoelastic driving fields, of the magnetoelastic response, and of the transmission nonreciprocity differ for both SAW modes.
Our experimental findings are well explained by a theoretical model based on a modified Landau--Lifshitz--Gilbert approach.
We show that the symmetries of the magnetoelastic response driven by Rayleigh- and shear horizontal SAWs complement each other, which makes it possible to excite spin waves for any relative orientation of magnetization and SAW propagation direction and, moreover, can be utilized to characterize surface strain components of unknown acoustic wave modes.

\end{abstract}

\maketitle

\section{Introduction}

Owing to the wealth of useful properties of surface acoustic waves (SAW) combined with the ease of launching and detecting SAWs on a piezoelectric crystal and low cost fabrication processes, SAW technology is employed in manifold ways in our daily life as rf-filters~\cite{Campbell.1998}, sensors~\cite{Lange.2008}, and lab-on-a-chip applications~\cite{Franke.2009}. 
However, basic research also benefited very profoundly from the use of SAWs, ranging from quantum phenomena in low-dimensional electron systems~\cite{A.Wixforth.1986} to acoustically operated nanophotonic devices~\cite{DanielA.Fuhrmann.2011}.

In recent years increasing attention has been paid to the coupling of SAWs with thin magnetic films. On the one hand, it was demonstrated that this coupling makes a new type of magnetic field sensors with an excellent signal-to-noise ratio possible~\cite{A.Kittmann.2018}. On the other hand, SAWs can excite spin waves (SW) in magnetic films, which turns out to be a fruitful playground for studying the SAW-SW coupling mechanism itself~\cite{Weiler.2011, Dreher.2012, Maekawa.1976, Xu.2020, Kobayashi.2017,Kurimune.2020}, characterizing the SW-dispersion relations~\cite{Gowtham.2015, M.Ku.2020} or even developing new kinds of "acoustic isolators"~\cite{Sasaki.2017, A.HernandezMinguez.2020, Tateno.2020, M.Ku.2020, Verba.2018, R.Verba.2019} based on nonreciprocity.

Although SAW propagation is in general reciprocal, i.e. invariant under inversion of the propagation direction, the coupling mechanism with the SW, and the SW propagation itself can be nonreciprocal. First, a pronounced nonreciprocal SW dispersion relation is obtained, inter alia, due to the interfacial Dzyaloshinskii--Moriya interaction (DMI) in a ferromagnetic/heavy metal bilayer~\cite{Verba.2018, M.Ku.2020}.
Secondly, the nonreciprocity of the SAW-SW coupling mechanism arises, because of a helicity mismatch between the magnetic driving fields, induced by the SAW and the fixed, right-handed rotational sense of the magnetic moments~\cite{Dreher.2012, Sasaki.2017, M.Ku.2020}.

Nevertheless, both the observation and possible technological application of these interesting effects are limited to certain experimental geometries, defined by the orientation of the static magnetization ${\bf M}$ with respect to the SW wave vector ${\bf k}_\text{SW}$, which is assumed to be determined by the wave vector of the SAW ${\bf k}_\text{SAW}={\bf k}_\text{SW}={\bf k}$~\cite{M.Ku.2020}.
This orientation dependence is caused by the SAW mode-specific symmetry of the magnetoelastic driving fields~\cite{Dreher.2012}.
So far, mainly Rayleigh-type (R) SAWs on piezoelectric LiNbO$_3$ substrates have been studied~\cite{Weiler.2011, Dreher.2012, Gowtham.2015, Xu.2020, M.Ku.2020, Sasaki.2017,Tateno.2020, Labanowski.2016}, which show vanishing magnetoelastic SW excitation efficiency for the often discussed backward volume magnetostatic SW mode (${\bf M} \parallel {\bf k}$) and magnetostatic surface SW mode (${\bf M} \perp {\bf k}$). 
Previous experiments, using shear horizontal (SH) SAW modes were not focused on resonant coupling of the SH-waves with SWs~\cite{Zhou.2014, A.Mazzamurro.2020,A.Kittmann.2018}.

In this study, we demonstrate in detail how the SAW mode-shape determines the symmetry of the magnetoelastic interaction and its nonreciprocal behavior, caused by the SAW-SW helicity mismatch effect.
Since we use a well-established 36$^\circ$-rotated Y-cut X-propagation LiTaO$_3$ substrate~\cite{Morgan.2007,Lange.2008}, which simultaneously supports both R- and SH-wave excitations, we can directly compare the symmetry of the magnetoelastic response of both SAW modes. Because the symmetry of the magnetoelastic driving fields of R- and SH-wave complement each other, efficient SW excitation is possible for any in-plane field geometry, which in fact could be a technologically relevant aspect.
In particular, the SH-wave allows efficient magnetoelastic coupling for ${\bf M} \parallel {\bf k}$ and ${\bf M} \perp {\bf k}$.
\begin{figure}
\includegraphics[width = 0.48\textwidth]{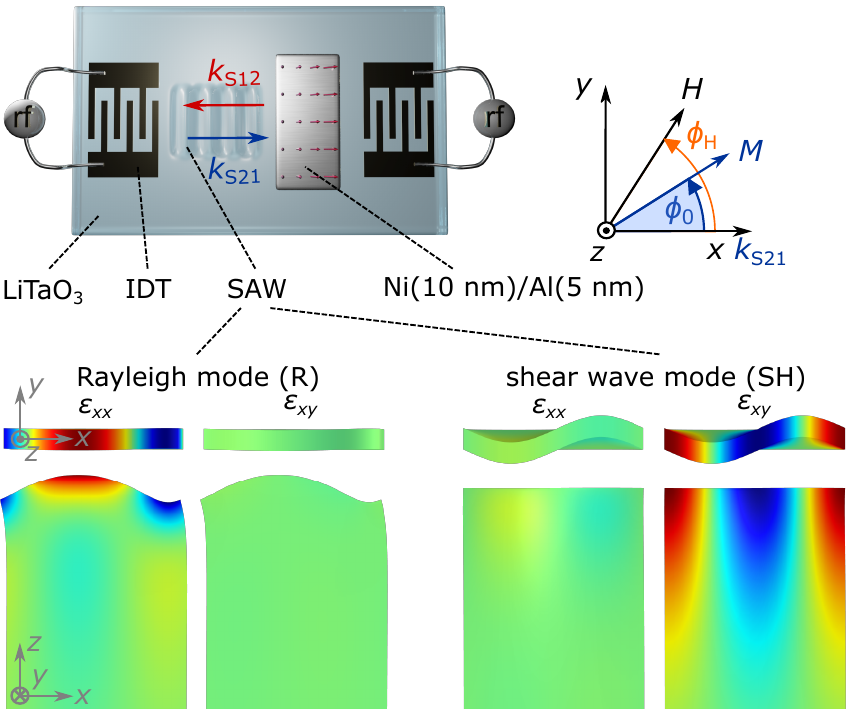}
\caption{
Schematic illustration of the experimental setup. R-wave and SH-wave modes can both be excited on the LiTaO$_3$ substrate. The FEM eigenfrequency simulation of the SAW modes shows the magnitude of the strain $\epsilon_{xx}, \epsilon_{xy}$ in false colors (green: low strain, blue and red: large negative and positive strain) and the exaggerated lattice deformation for the LiTaO$_3$/Ni(10~nm)/Al(5~nm) layer stack and a wavelength of 1.17~$\mu$m.
The 1000~$\mu$m long Ni film is placed centered between the 1600~$\mu$m distant IDTs. The nonreciprocal behavior of both SAW modes is characterized by different transmission amplitudes $\Delta S_{21}$ and $\Delta S_{12}$ for oppositely propagating SAWs $k_{S21}$ and $k_{S12}$. 
\label{fig:1}
}
\end{figure}

\section{Theory and FEM simulation}

First, we discuss the nonreciprocal transmission characteristics of the magnetoacoustic sample displayed in Fig.~\ref{fig:1}.
A SAW is excited on the piezoelectric substrate, once an alternating voltage with the resonance frequency $f = c_\mathrm{SAW} / \lambda$ of the interdigital transducer (IDT) is applied. Here, $c_\mathrm{SAW}$ is the propagation velocity of the SAW and $\lambda$ is the wavelength, given by the periodicity of the IDT.
In this study, we use a 36$^\circ$-rotated Y-cut X-propagation LiTaO$_3$ substrate, which has been extensively exploited for building high-frequency bandpass filters~\cite{Morgan.2007} and for biosensing applications~\cite{Lange.2008}. 
This substrate supports predominantly a SH-mode ($c_\mathrm{SAW}=4112$~m/s), but additionally a R-mode ($c_\mathrm{SAW}=3232$~m/s)~\cite{Nakamura.1977} resulting in different IDT resonance frequencies. 
The lattice displacement of both modes is depicted in Fig.~\ref{fig:1} for the LiTaO$_3$/Ni(10~nm)/Al(5~nm) layer stack.
Depending on the wave mode, either SAW will induce specific strain components in the thin ferromagnetic Ni-film and dynamically modulate the magnetic free energy due to inverse magnetostriction. 
Because of the high magnetoelastic coupling efficiency of Ni, we here neglect non-magnetoelastic interaction, like magneto-rotation coupling~\cite{Maekawa.1976, Xu.2020, M.Ku.2020}, spin-rotation coupling~\cite{Matsuo.2011, Matsuo.2013, Kobayashi.2017} or gyromagnetic coupling \cite{Kurimune.2020}.

The SAW-SW interaction can be described by dynamic magnetoelastic driving fields, which exert a torque on the static magnetization ${\bf M}$~\cite{Weiler.2011}. The resulting attenuated precession of ${\bf M}$ is then given by the Landau--Lifshitz--Gilbert equation. 
As shown in Fig.~\ref{fig:1}, an external magnetic field ${\bf H}$ with the direction $\phi_H$ is applied to align the static magnetization ${\bf M}$ to the angle $\phi_0$ in the film plane.
According to Ref.~\cite{M.Ku.2020}, the magnetoelastic driving fields ${\bf h}(x,t)$ with the normalized out-of-plane component $\tilde{h}_1$ and in-plane component $\tilde{h}_2$, both being perpendicular to ${\bf M}$, are a function of the SAW power $P_\mathrm{SAW}$
\begin{equation}
 {\bf h}(x,t) =
    \begin{pmatrix}
        \tilde{h}_1 \\
        \tilde{h}_2
    \end{pmatrix}
 \sqrt{ \frac{k^2}{R\, \omega W}} \sqrt{P_\text{SAW}(x)}~\text{e}^{i(kx-\omega t)}.
 \label{eq:1}
\end{equation}
Here, $k$ and $\omega$ are the wave vector and angular frequency of the SAW, respectively. $W$ is the aperture of the IDT and $R$ is a constant factor, depending on the type of the SAW mode.
Following Dreher \textit{et al.}~\cite{Dreher.2012}, the symmetry of the normalized magnetoelastic driving fields $\tilde{h}_1$ and $\tilde{h}_2$ for vanishing strain $\epsilon_{yy}$ are
\begin{align}
\label{eq:2}
    \begin{pmatrix}
        \tilde{h}_1 \\
        \tilde{h}_2
    \end{pmatrix}
 &=
    \begin{pmatrix}
        \tilde{h}_1^\mathrm{Re} + i \tilde{h}_1^\mathrm{Im} \\
        \tilde{h}_2^\mathrm{Re} + i \tilde{h}_2^\mathrm{Im}
    \end{pmatrix}
\nonumber \\
&=
\frac{2}{\mu_0}
    \begin{pmatrix}
        b_{xz} \cos \phi_0 +  b_{yz} \sin \phi_0\\
        b_{xx} \sin \phi_0 \cos \phi_0 - b_{xy} \cos (2\phi_0)
    \end{pmatrix}.
\end{align}
The magnetoelastic parameters are $b_{kl} = b_1 \tilde{a}_{kl}$ ($kl \in \{ xx, xy, xz, yz$\}) with an isotropic magnetoelastic coupling constant $b_1=b_2$ for polycrystalline films. 

\begin{figure}[h]
\includegraphics[width = 0.48\textwidth]{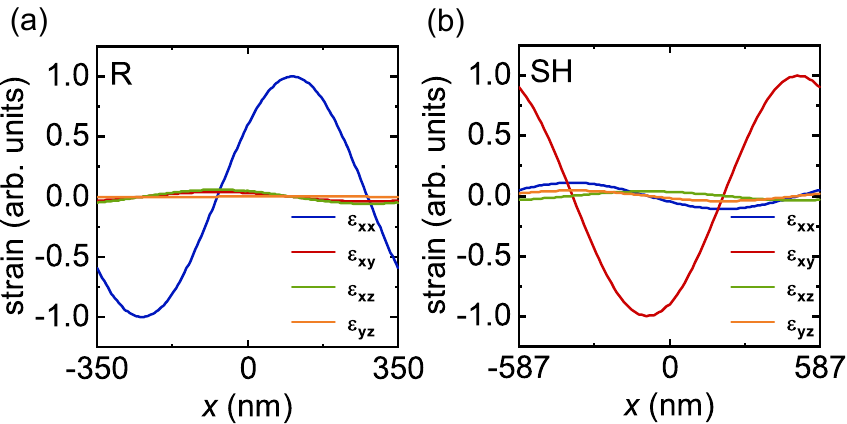}
\caption{
\label{fig:2}
FEM eigenfrequency simulation of the LiTaO$_3$/Ni(10~nm)/Al(5~nm) structure, carried out with Comsol~\cite{Comsol.} to determine the phase and normalized magnitude of the relevant strain components for magnetoelastic SAW-SW interaction. The lengths of the simulation geometries that correspond to one wavelength $\lambda$ of the (a) R-wave and (b) SH-wave are adjusted to match the resonance frequencies of the experiment.
The thin film parameters that are used for the FEM simulation for Ni (Al) are: density $\rho=8900 ~ \mathrm{kg}/\mathrm{m}^3$~\cite{Mills.1999} (2700~$\mathrm{kg} / \mathrm{m}^3$~\cite{Comte.2002}), Young's modulus $E = 218$~GPa~\cite{Ledbetter.1973} (70~GPa~\cite{Comte.2002}), and Poisson's ratio $\nu = 0.3$~\cite{Ledbetter.1973} (0.33~\cite{Comte.2002}).
The parameters for the anisotropic single-crystal LiTaO$_3$ are taken from Ref.~\cite{Comsol.}.
}
\end{figure}

The complex amplitudes of the normalized strain $\tilde{a}_{kl} = \epsilon_{kl,0} / (|k| |u_{z,0}|)$, where $u_{z,0}$ is the amplitude of the lattice displacement in the $z$ direction, can be determined by a finite element method (FEM) simulation~\cite{M.Ku.2020}. 
Results and parameters of the FEM simulation are given in Table~\ref{tab:table1}.
In Figs.~\ref{fig:2}(a) and \ref{fig:2}(b) we show the calculated strain components of the R- and SH-wave in the center plane of the Ni-film, as simulated for the LiTaO$_3$/Ni(10~nm)/Al(5~nm) structure and for the resonance frequencies of the IDTs. Because the longitudinal strain $\epsilon_{xx}$ is dominating in the R-mode, the main symmetry of the driving field in Eq.~\eqref{eq:2} is $\propto \sin \phi_0 \cos \phi_0$~\cite{Weiler.2011, Dreher.2012, Gowtham.2015}. In contrast, $\epsilon_{xy}$ is dominating for the SH-mode and the expected symmetry of the main driving field component is $\propto \cos (2 \phi_0)$. We thus expect qualitatively different dependencies of SAW absorption on the magnetization direction $\phi_0$ for magnetoacoustic resonance driven by R and SH SAWs.

Smaller strain components potentially cause nonreciprocal SAW transmission due to the SAW-SW helicity mismatch effect~\cite{Dreher.2012, Sasaki.2017, M.Ku.2020}.
For the R-wave (SH-wave) the strain components $\epsilon_{xy,xz}$ ($\epsilon_{xx,yz}$) are phase shifted by $+90^\circ$  $(-90^\circ)$ with respect to the main strain component $\epsilon_{xx}$ ($\epsilon_{xy}$). Therefore, the corresponding amplitudes of $\epsilon_{kl,0}$, $\tilde{a}_{kl}$, and $\tilde{h}_{1,2}$ are complex and we can separate the real and imaginary parts of $\tilde{h}_{1,2}$ with $\tilde{h}_{1,2}^\mathrm{Re} \equiv \mathrm{Re}(\tilde{h}_{1,2})$ and $\tilde{h}_{1,2}^\mathrm{Im} \equiv \mathrm{Im}(\tilde{h}_{1,2})$.
By reversing the propagation direction of the SAW ($k_{S21} \rightarrow k_{S12}$), the phase difference between the complex and the main strain components becomes inverted. Thus, the helicity of the driving fields changes. This is expressed by the inversion of the sign of the complex $\tilde{a}_{kl}$ in Table~\ref{tab:table1}. 
In combination with the fixed, right-handed rotational sense of the magnetization precession, the SAW-SW helicity mismatch effect arises, inducing nonreciprocal efficiency of SW excitation and SAW absorption.

We now expand the theory, as outlined in Ref.~\cite{M.Ku.2020}, in terms of generalized driving field components of Eq.~\eqref{eq:2}.
This model is based on the "Landau--Lifshitz--Gilbert approach" of Ref.~\cite{Dreher.2012} and thus considers implicitly the backaction of the precessing magnetization on the SAW for low excitation amplitudes ~\cite{Dreher.2012}.
For zero Dzyaloshinskii--Moriya interaction, the absorbed power of the SAW, which is used to drive magnetization precession, is expressed by
\begin{widetext}
\begin{align}
P_\text{abs} = P_0
\Biggl[
1 - \text{exp} &
\Biggl(
- C M_s
\frac{\alpha H_\omega}
{
\left[
 H_\omega ^2 (1 + \alpha^2)
- H_{11} H_{22}
 \right]^2 
 + \left[ 
 \alpha H_\omega (H_{11} + H_{22})
 \right]^2
 } 
 \nonumber \\
 \times \bigg\{ 
&
\left[
H_\omega ^2 (1 + \alpha^2) + H_{11}^2
\right]
\Big[ \left(\tilde{h}_2^\mathrm{Re}\right)^2 + \left(\tilde{h}_2^\mathrm{Im}\right)^2 \Big]
 \nonumber\\
 + 
 &
 \left[
 H_\omega (H_{11} + H_{22})
 \right]
\left[2 \left( 
\tilde{h}_1^\mathrm{Re} \tilde{h}_2^\mathrm{Im} - \tilde{h}_1^\mathrm{Im} \tilde{h}_2^\mathrm{Re} 
\right) \right]
  \nonumber\\
 +
 &
 \left[
 H_\omega ^2 (1 + \alpha^2) + H_{22}^2
\right]
\Big[\left(\tilde{h}_1^\mathrm{Re}\right)^2 + \left(\tilde{h}_1^\mathrm{Im}\right)^2 \Big]
 \bigg\} \Biggl) \Biggl].
 \label{eq:3}
\end{align}
\end{widetext}
With respect to the initial power $P_0$, the power of the traveling SAW is exponentially decaying while propagating through the magnetic thin film. The decay rate depends on the effective SW damping constant $\alpha$ and $C, H_\omega, H_{11}$, $H_{22}$, which are defined in Ref.~\cite{M.Ku.2020}.
Eq.~\eqref{eq:3} is derived by taking into account, (i) the Zeeman energy, (ii) a uniaxial in-plane magnetic anisotropy field $H_\mathrm{ani}$, under an angle $\phi_\mathrm{ani}$ with the $x$ axis, (iii) an out-of-plane magnetic anisotropy field $H_k$, counteracting the magnetic shape anisotropy, (iv) the dipolar fields of the SW~\cite{Kalinikos.}, (v) the magnetic exchange exchange interaction, and (vi) the magnetoelastic driving fields of Eq.~\eqref{eq:1}.

Finally, to directly fit the exponent of Eq.~\eqref{eq:3} to the experimentally determined relative change of the SAW transmission $\Delta S_{ij}$ on the logarithmic scale, the fit equation is given by
\begin{equation}
\Delta S_{ij}
= 10 \lg \left( \frac{P_0 - P_\text{abs}}{P_0} \right).
\label{eq:4}
\end{equation}
The symmetry of $\Delta S_{ij}$ is determined by the symmetry of the driving fields, as discussed before. We obtain, for the main symmetry of R-waves (SH-waves), $\Delta S_{ij} \propto (\sin \phi_0 \cos \phi_0)^2$ ($\Delta S_{ij} \propto \sin^2 2 \phi_0$). Employing the FEM study, the real and imaginary terms in Eq.~\eqref{eq:2} can be identified. For R-waves and SH-waves, the expected leading term that causes nonreciprocity ($\Delta S_{21}-\Delta S_{12} \neq 0$) is $\propto \tilde{h}_1^\mathrm{Im} \tilde{h}_2^\mathrm{Re}$.
This nonreciprocity for R-waves (SH-waves) is mediated by the strain component $\epsilon_{xz}$ ($\epsilon_{yz}$) with the symmetry of the nonreciprocity being proportional to $\sin \phi_0 \cos^2 \phi_0$ ($\sin \phi_0 \cos 2 \phi_0$).
\begin{table}
\caption{\label{tab:table1}
Results of the FEM simulation.
The normalized complex amplitudes of the strain tensor are $\tilde{a}_{kl} = \epsilon_{kl,0} / (|u_{z,0}| |k|) $ with $kl \in \{ xx, xy, xz, yz$\}.
The errors are assumed to be of the order of $\pm10\%$ of $\tilde{a}_{xx}$ ($\tilde{a}_{xy}$) for the R-wave (SH-wave).
}
\begin{ruledtabular}
\begin{tabular}{ccccccc}
  & $f$~(GHz) & $c_\mathrm{SAW}$~(m/s) & $\tilde{a}_{xx}$ & $\tilde{a}_{xy}$ & $\tilde{a}_{xz}$ & $\tilde{a}_{yz}$   \\
\colrule
R  & 4.47 & 3105 & 0.613 & $\pm i$0.024 & $\pm i$0.037 & 0\\
SH & 3.47 & 4075 & $\mp i$0.53 & 4.85 & -0.18 & $\mp i$0.21\\
\end{tabular}
\end{ruledtabular}
\end{table}

\section{Experimental methods}

To prove the theoretical predictions for R- and SH-waves, we fabricate a magnetic thin film sample, as depicted in Fig.~\ref{fig:1}.
IDTs with a periodicity of 3.4~$\mu$m and a 200~$\mu$m aperture are e-beam lithographically defined on a 36$^\circ$-rotated Y-cut X-propagation LiTaO$_3$ substrate, evaporating 5~nm Ti and 70~nm of Al. 
The rectangular-shaped Ni(10~nm)/Al(5~nm) film is deposited by dc magnetron sputterdeposition (base pressure $< 10^{-8}$~mbar) at room temperature and positioned in the middle between the two 1600~$\mu$m spaced IDTs. The Ar sputter pressure is kept constant at $3.5 \times 10^{-3}$~mbar and the sample holder is rotated during sputtering.

We carried out superconducting quantum interference device-vibrating sample magnetometry (SQUID-VSM) measurements to determine the saturation magnetization ($M_s = 408$~kA/m).
Additionally, broadband ferromagnetic resonance (FMR) measurements are performed to obtain values for the g-factor, the out-of-plane magnetic anisotropy $H^{\mathrm{FMR}}_k$, and the effective damping constant $\alpha_\mathrm{eff}^\mathrm{FMR}= \mu_0 \Delta H \gamma / (2 \omega)+ \alpha^\mathrm{FMR}$~\cite{M.Ku.2020}, which includes Gilbert damping $\alpha^\mathrm{FMR}$ and inhomogeneous line broadening $\Delta H$.

To characterize the delayline sample, and to measure the magnitude of the complex transmission signal, $S_{ij}$  with $ij \in \{ 21, 12 $\} we employ standard network analyzer measurements~\cite{Ekstrom.2017,Hiebel.2011}. 
Nonreciprocal effects are studied by comparing the $S_{ij}$ obtained for oppositely propagating SAWs with $k_{S21}$ and $k_{S12}$.

\section{Discussion}

\begin{figure}[h]
\includegraphics[width = 0.48\textwidth]{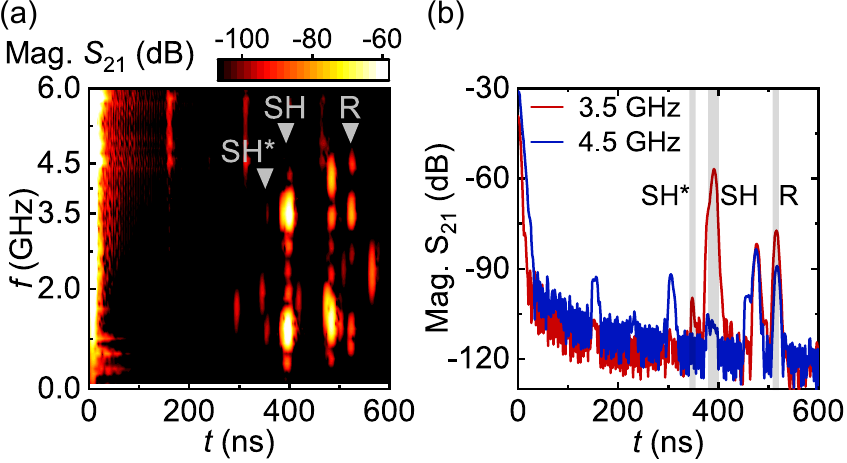}
\caption{
\label{fig:3}
(a) Various acoustic wave modes are visible in the $S_{21}(t, f, \mu_0 H = -200~\mathrm{mT})$ spectrogram. Signal components, that we identified are (i) electromagnetic crosstalk  at about $0$~ns, (ii) SH*-mode at 349~ns (discussed later), (iii) two harmonic resonances of the SH-mode at 390~ns, (iv) bulk waves that are multiple times reflected on the upper and lower side of the LiTaO$_3$ substrate at 475~ns, and (v) three harmonic resonances of the R-mode and additionally bulk waves at 515~ns.
(b) Line cuts of (a) at 3.5~GHz and 4.5~GHz, which correspond to the 3rd and 5th harmonic resonance frequencies of the SH- and R-mode. The adjusted time gates for the SAW modes are depicted in gray. The peak at 515~ns and 3.5~GHz does not show a magnetoelastic response and is thus attributed to bulk waves.
%
}
\end{figure}

The acoustic wave transmission magnitude $S_{21}$ in the time-domain as a function of frequency is characterized in Fig.~\ref{fig:3}(a) at a quite high magnetic field of $-200$~mT and, therefore, far off the SW resonance.
The obtained spectrogram contains electromagnetic crosstalk at $t \approx 0$, acoustic bulk waves, SAWs, and also some higher harmonic resonances, as described in more detail in the caption of Fig.~\ref{fig:3}.
By comparing the SAW propagation velocities $c_\mathrm{SAW} = 1600~\mathrm{\mu m} / t$ with the results from the FEM simulation given in Table~\ref{tab:table1}, we identify the R-mode at 515~ns ($3107$~m/s) and the SH-mode at 390~ns ($4103$~m/s).

Now, we turn to the detailed study of the symmetry of the magnetoacoustic response and its nonreciprocity for both different SAW modes.
To do so, we use adjusted time gates for both modes, as depicted in Fig.~\ref{fig:3}(b). Then we apply inverse Fourier transformation to solely measure the peak transmission of each individual SAW mode in the frequency domain at 4.5~GHz for the R-mode and at 3.5~GHz for the SH-mode.
The relative change of the background-corrected SAW transmission magnitude, which is caused by SAW-SW interaction is defined as $\Delta S_{ij} (\mu_0 H) = S_{ij} (\mu_0 H) - S_{ij} (- 200~\mathrm{mT})$.

\begin{figure}[h]
\includegraphics[width = 0.48\textwidth]{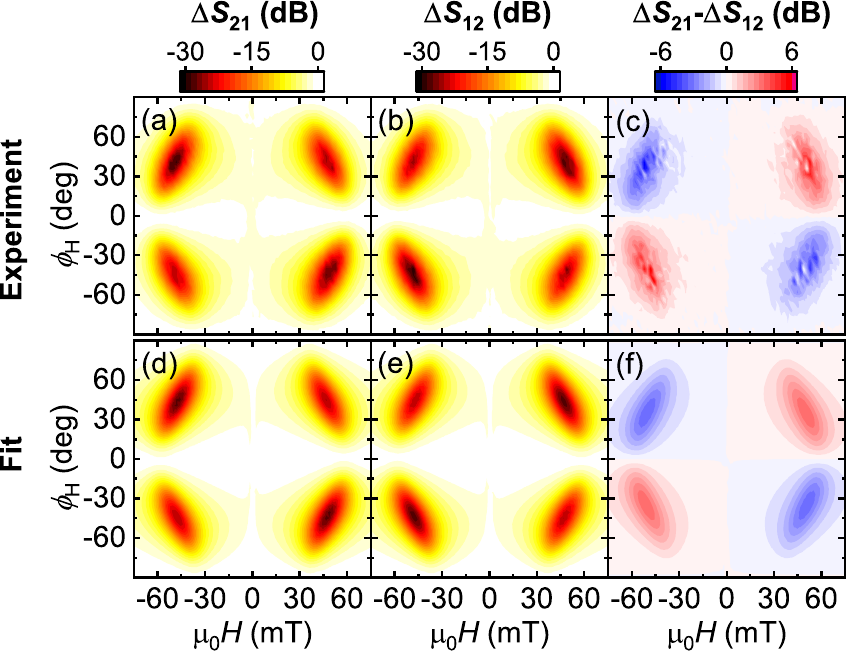}
\caption{
\label{fig:4}
Transmission nonreciprocity of the R-wave at 4.47~GHz. The experimental data $\Delta S_{21}$ (a) and $\Delta S_{12}$ (b) demonstrate the expected four-fold symmetry, caused by dominant longitudinal strain  $\epsilon_{xx}$. Additionally, the nonreciprocal behavior $\Delta S_{21}-\Delta S_{12}$ in (c) is four-fold and induced by the strain component $\epsilon_{xz}$.
Experiment and fit (d)-(f) show excellent agreement.
}
\end{figure}

In Figs.~\ref{fig:4}(a) and \ref{fig:4}(b) we show $\Delta S_{ij}$ for the R-mode as a function of the external magnetic field magnitude $H$ and direction $\phi_H$. Since the resonance fields $H_\mathrm{res}$ are much higher \mbox{($>30$~mT)} than the uniaxial in-plane anisotropy (1.4~mT, fit results in \mbox{Table}~\ref{tab:table2}), $\mathbf{M}$ and $\mathbf{H}$ are approximately parallel (\mbox{$\phi_0 \approx \phi_H$}) for $H_\mathrm{res}$ and the symmetry of the main driving field shows up in Figs.~\ref{fig:4}(a) and \ref{fig:4}(b).
As expected from theory, we observe the fourfold symmetry $\Delta S_{ij} \propto (\sin \phi_0 \cos \phi_0)^2$ for the R-wave~\cite{Weiler.2011, Dreher.2012, Gowtham.2015}.

Eqs.~\eqref{eq:1}-\eqref{eq:4} are used to fit the experimental results of Figs.~\ref{fig:4}(a) and \ref{fig:4}(b), following the curve fitting procedure described in Ref.~\cite{M.Ku.2020}. 
Because we do not know the parameter $R$ in Eq.~\eqref{eq:1}, the fitting parameters are $\alpha$, $H_k$, $b_{kl} / \sqrt{R}$ with $kl = xx, xy, xz, yz$, and the uniaxial in-plane magnetic anisotropy.
The fits in Figs.~\ref{fig:4}(d) and \ref{fig:4}(e) show excellent agreement with the experiment. Furthermore, the fit results, as summarized in Table~\ref{tab:table2}, are in accordance with the FMR data for $H_k^\mathrm{FMR}$ and $\alpha_\mathrm{eff}^\mathrm{FMR}$. Note that the FMR experiments were performed on reference samples (same sputterdeposition run as SAW samples) 20 months after the SAW measurements had been carried out, explaining slight deviations due to possible degeneration of the Ni thin film.
\begin{table*}
\caption{\label{tab:table2}
Summary of the results obtained by fitting the SAW transmission $\Delta S_{21}$ of Figs.~\ref{fig:4}(a), \ref{fig:5}(a), and of the FMR measurements. Additional parameters that are used for the fit are $M_s = 408$~kA/m (obtained by SQUID-VSM), $\gamma = 193.5 \times 10^{9}$~rad/(s T) (obtained by FMR), and the exchange constant $A=7.7$~pJ/m~\cite{Wilts.1972}. Further fit results are the direction $\phi_\mathrm{ani} = (83.6 \pm 3.6) ^\circ$ and magnitude $\mu_0 H_\mathrm{ani} = (1.4 \pm 1)$~mT of the in-plane uniaxial anisotropy easy axis.
}
\begin{ruledtabular}
\begin{tabular}{cccccccccc}
  & & \multicolumn{6}{c}{SAW}&\multicolumn{2}{c}{FMR}\\
  & $f$~(GHz) & $H_k~\left(\frac{\mathrm{kA}}{\mathrm{m}} \right)$ & $\alpha \left( 10^{-3}\right)$ & 
  $\frac{b_{xx}}{\sqrt{R}}$~$\left( \frac{\mu \mathrm{T}}{\sqrt{\frac{\mathrm{J}}{\mathrm{m}^3}}} \right)$ & 
  $\frac{b_{xy}}{\sqrt{R}}$~$\left( \frac{\mu \mathrm{T}}{\sqrt{\frac{\mathrm{J}}{\mathrm{m}^3}}} \right)$ & 
  $\frac{b_{xz}}{\sqrt{R}}$~$\left( \frac{\mu \mathrm{T}}{\sqrt{\frac{\mathrm{J}}{\mathrm{m}^3}}} \right)$ & 
  $\frac{b_{yz}}{\sqrt{R}}$~$\left( \frac{\mu \mathrm{T}}{\sqrt{\frac{\mathrm{J}}{\mathrm{m}^3}}} \right)$ & $H_k^\mathrm{FMR}~\left(\frac{\mathrm{kA}}{\mathrm{m}} \right)$ & $\alpha_\mathrm{eff}^\mathrm{FMR} \left( 10^{-3}\right)$  \\
\colrule
R  & 4.47 & 158.2$\pm$0.1 & 69$\pm$1 & 20.80$\pm$0.02 & $+ i$(1.68$\pm$0.04) & $+ i$(1.03$\pm$0.02) & 0.03$\pm$0.11 & 127.4$\pm$0.2 & 75$\pm$4\\
SH & 3.47 & 161.7$\pm$0.1 & 76$\pm$2 & $- i$(5.64$\pm$0.13) & 15.23$\pm$0.01 & -(0.06$\pm$0.23) & $- i$(0.55$\pm$0.03) & 127.4$\pm$0.2 & 87$\pm$4\\
\end{tabular}
\end{ruledtabular}
\end{table*}

The experimentally determined symmetry of the nonreciprocity $\Delta S_{21} - \Delta S_{12}$ of the R-mode is depicted in Fig.~\ref{fig:4}(c). As expected from theory, the nonreciprocity is caused by the vertical shear strain $\epsilon_{xz}$ and is proportional to $\cos^2(\phi_0) \sin(\phi_0)$. Excellent agreement between Fig.~\ref{fig:4}(c) and Fig.~\ref{fig:4}(f), which is obtained by subtracting the fit curves of Fig.~\ref{fig:4}(d) and Fig.~\ref{fig:4}(e), further validates the theoretical model.

\begin{figure}[h]
\includegraphics[width = 0.48\textwidth]{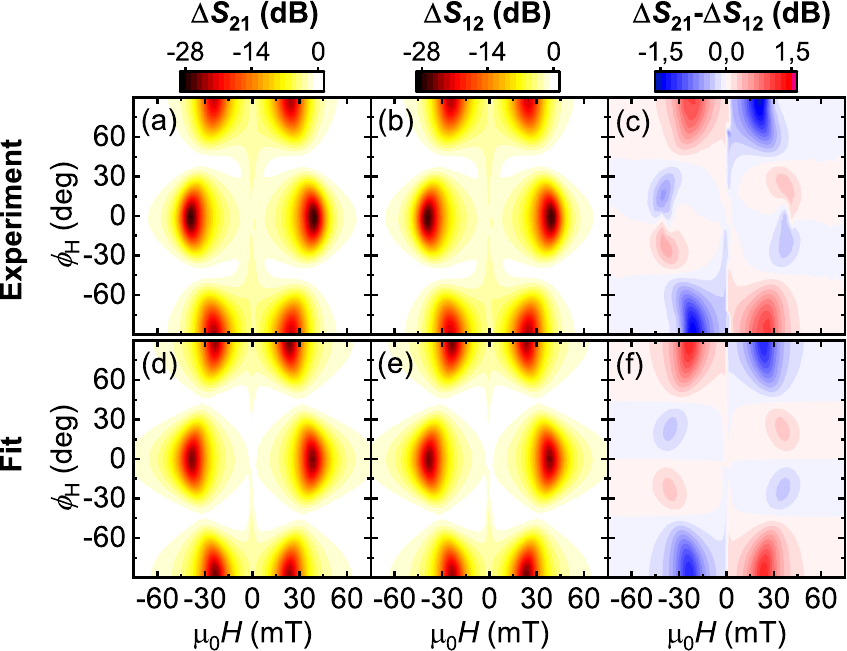}
\caption{
\label{fig:5}
Transmission nonreciprocity of the SH-wave at 3.47~GHz. The experimental data $\Delta S_{21}$ (a) and $\Delta S_{12}$ (b) show the expected symmetry for SH-waves with the dominant strain $\epsilon_{xy}$. A complicated nonreciprocal behavior is observed in (c), which is attributed to the nonzero strain component $\epsilon_{yz}$.
Experiment and fit (d)-(f) show excellent agreement.
}
\end{figure}

The results for the SH-wave are shown in Fig.~\ref{fig:5}. Since the main strain component of the SH-wave $\epsilon_{xy}$ induces driving fields with a symmetry proportional to $\cos (2 \phi_0)$, the experimental response $\Delta S_{ij}$ in Figs.~\ref{fig:5}(a) and \ref{fig:5}(b) differs, but complements the symmetry of the R-wave.
The fit results in Figs.~\ref{fig:5}(d) and \ref{fig:5}(e) reproduce the experiment again very well. Moreover, the fit parameters $H_k$ and the effective damping $\alpha$, which depends on the SAW frequency, are in good agreement with the fit parameters of the R-wave and with the FMR measurements given in Table~\ref{tab:table2}.

The experimentally determined nonreciprocity of the SH-wave in Fig.~\ref{fig:5}(c) has a different symmetry with a lower magnitude than the R-wave. As expected from theory, the strain $\epsilon_{yz}$ causes the SAW-SW helicity mismatch effect with the symmetry being proportional to $\sin \phi_0 \cos (2 \phi_0)$, as observed in the experiment. Again, the difference of the fits $\Delta S_{21} - \Delta S_{12}$ in Fig.~\ref{fig:5}(f) agrees well with the nonreciprocity of the experiment, also confirming the theoretical model for SH-waves.

An approximate estimation of the magnitude of the dominant strain component gives $\epsilon_{xx,0} \approx 12 \times 10^{-6}$ for the R-mode and $\epsilon_{xy,0} \approx 23 \times 10^{-6}$ for the SH-mode~\cite{EstimateStrain.0001}.
\nocite{Croset.2017}
With these value we estimate the in-plane magnetization precession amplitude to be of the order of $\phi_\text{M,ip} \approx 0.6^\circ$ for the R-mode and $\phi_\text{M,ip} \approx 3^\circ$ for the SH-mode~\cite{OtherRef.0002}, which is one order of magnitude lower as compared to Ref.~\cite{Casals.2020}.
For both SAW modes, the transmission $\Delta S_{21}$ does not change with the output power of the vector network analyzer $P_\text{VNA}$ in the studied range of -15~dBm to +15~dBm. Thus, SAW propagation and SW excitation of the presented experiments ($P_\text{VNA}=15$~dBm) are in the linear regime.

So far, we have presented the results for the 5th and 3rd harmonic resonance frequency of the R- and SH-mode. We perform similar measurements for all transmission peaks visible in Fig.~\ref{fig:3}(a). Since only surface modes are expected to induce considerable magnetoelastic driving fields, we make use of this assumption and identify the surface modes by looking at the magnitude of the absorbed SAW power, caused by the SAW-SW interaction $\Delta S_{ij} (\mu_0 H, \phi_H)$.
None of the other transmission peaks in Fig.~\ref{fig:3}(a) shows a magnetoelastic response, except the weak but still detectable SAW mode at $t=349$~ns. 
This mode cannot be identified from either a literature search~\cite{Nakamura.1977} or our FEM eigenfrequency simulation based on its propagation velocity. We still name this mode SH*-mode.

\begin{figure}[h]
\includegraphics[width = 0.48\textwidth]{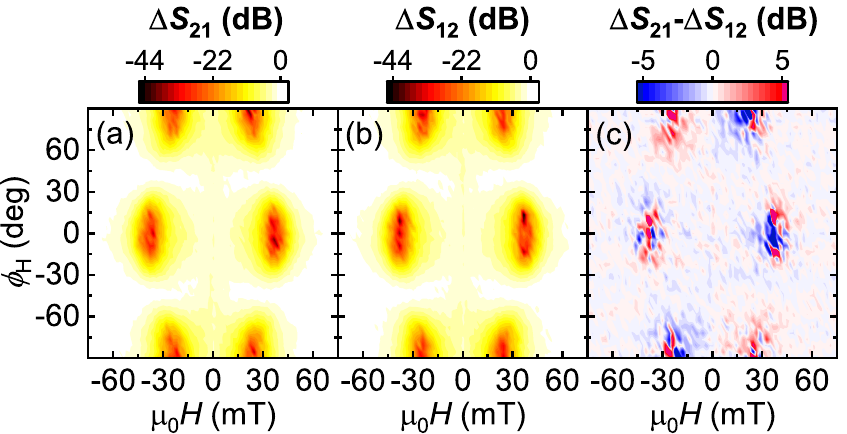}
\caption{
\label{fig:6}
Experimental results for the transmission nonreciprocity of the SH*-wave at 3.52~GHz, revealing a similar symmetry as the SH-wave response shown in Fig.~\ref{fig:5}.
}
\end{figure}

Given the example of this SH*-wave mode, we demonstrate that an unknown SAW mode can also be characterized in terms of its strain components by employing SAW driven SW spectroscopy.
The magnetoelastic response $\Delta S_{ij}$ of this SH*-mode is depicted in Fig.~\ref{fig:6}. Because the symmetry of $\Delta S_{ij}$ is an unambiguous indication of the strain component $\epsilon_{xy}$ in Eq.~\eqref{eq:2}, we conclude and experimentally confirm that this mode must be a shear horizontal type wave. 
The smaller strain components, which are phase shifted with respect to the main strain component, show up in the nonreciprocal response in Fig.~\ref{fig:6}(c). Despite low signal-to-noise ratio, the nonreciprocity of the SH*-wave agrees with the nonreciprocity of the SH-wave, caused by the phase shifted strain $\epsilon_{yz}$.
We infer that the SH*-mode is also a surface acoustic wave with low transmission and strain tensor elements, which indicate it to be similar to the SH-mode.

To find out more details about the SH*-mode, we carried out additional time-dependent FEM simulations, using the exact geometry of the LiTaO$_3$ sample. These simulations include, in contrast to the eigenfrequency FEM simulation, acoustic wave reflections, and secondary induced acoustic waves due to electromagnetic crosstalk.
The time-dependent simulation shows that electromagnetic crosstalk causes low-amplitude secondary acoustic wave excitation at the edge of the magnetic film, close to the exciting IDT. Due to the reduced propagation path, the propagation time of this secondary mode is lowered by about $300~\mu\text{m} / (4075~\mathrm{m/s}) = 74$~ns in comparison to the SH-mode. This, however, does not agree with the experimental findings for the SH*-mode (time delay of 42~ns in Fig.~\ref{fig:3}). Since the propagation time of none of the acoustic wave modes of
the simulation matches with the propagation time of the SH*-mode of $349$~ns, we have to conclude that we can not reproduce the SH*-mode in the FEM simulations.
Furthermore, because we observe the SH*-mode in the off-resonant condition in Fig.~\ref{fig:3}, this mode cannot be a secondary elastic wave generated by the backaction of the  magnetization precession, as described in Ref.~\cite{Azovtsev.2019}. 

\begin{figure}[h]
\includegraphics[width = 0.48\textwidth]{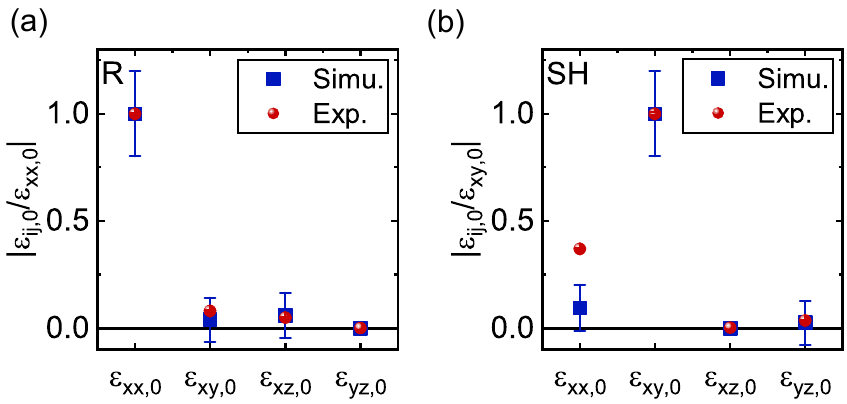}
\caption{
\label{fig:7}
Comparison of the experiment results and FEM simulation of the normalized strain components for the (a) R-wave and (b) SH-wave.
}
\end{figure}

Finally, we discuss the magnitude of the magnetoelastic driving fields of the fits $b_{kl} / \sqrt{R}$ (Table~\ref{tab:table2}) by comparing these values with $\tilde{a}_{kl,0}$ of the FEM simulation (Table~\ref{tab:table1}).
Thus, the normalized strain components of the fit $\frac{\epsilon_{kl,0}}{\epsilon_{xx,0}}  =  \frac{b_{kl,0}}{b_{xx,0}}$ and of the simulation $\frac{\epsilon_{kl,0}}{\epsilon_{xx,0}}  =  \frac{\tilde{a}_{kl,0}}{\tilde{a}_{xx,0}}$ are shown for the R-wave and SH-wave (normalized to $\epsilon_{xy,0}$) in Figs.~\ref{fig:7}(a) and \ref{fig:7}(b), respectively.
Note that the sign of $\tilde{a}_{kl,0}$ of the simulation is in accordance with the experimental result for $b_{kl,0}$ and we plot absolute values.

The excellent agreement, except for $\epsilon_{xx,0}$ (SH), between FEM simulation and experimental results again confirms the theory.
Due to additional, non-magnetoelastic coupling mechanisms like magneto-rotation coupling~\cite{Maekawa.1976, Xu.2020, M.Ku.2020} or spin-rotation coupling~\cite{Matsuo.2011, Matsuo.2013, Kobayashi.2017}, corrections for the driving fields are potentially the reason for the deviation of $\epsilon_{xx,0}$ of the SH-wave. 
Furthermore, we assume in the theory section that the SAW mode shape is fixed [Eq.~\eqref{eq:1}], but only the amplitudes decay exponentially with increasing SAW propagation in the magnetic film [Eq.~\eqref{eq:3}]. Because the shape of the acoustic wave transforms from a SH-wave with an extremely large penetration depth in $z$ direction at the bare LiTaO$_3$ surface to a SH-wave with much shorter penetration depth at the shorted LiTaO$_3$/Ni(10~nm)/Al(5~nm) surface~\cite{Nakamura.1977}, this assumption holds only for long propagation distances through the magnetic film and also explains deviations between the FEM eigenfrequency study and the experiment.

\section{Conclusion}

In conclusion, we have extended the theoretical model of SAW driven SW spectroscopy~\cite{M.Ku.2020} in terms of arbitrary magnetoelastic driving fields. This model now describes the experimental results for both R- and SH-wave transmissions on a LiTaO$_3$/Ni(10~nm)/Al(5~nm) sample in an excellent way. Additionally, the fit values were cross-checked with FMR measurements and FEM simulations, reasonably confirming the theoretical model. The SAW-SW helicity mismatch effect causes nonreciprocal SAW transmission for the SH- and R-modes, with symmetries proportional to $\sin \phi_0 \cos (2 \phi_0)$ and $\sin \phi_0 \cos^2 (\phi_0)$, respectively.
Since the symmetry of the induced magnetoelastic driving fields of all relevant strain components is ambiguous, the SAW type can in general be identified by studying the symmetry behavior of the magnetoelastic response.

Furthermore, the symmetry of the magnetoelastic driving fields of R- and SH-waves complement each other. Thus, a large SAW-SW interaction is observed for any in-plane field geometry on the LiTaO$_3$ sample.
This is especially interesting because processing and transport of data with SWs in magnonics are usually carried out in either the ${\bf M} \perp {\bf k}$ or the ${\bf M} \parallel {\bf k}$ geometry~\cite{AASerga.2010}.
So far, it has not been possible to excite SWs with Rayleigh waves in these geometries in an efficient way. As we demonstrate in this study, efficient magnetoacoustic excitation of SWs in exactly these geometries is achieved by using SH-waves. 
This could be additionally used to excite backward volume magnetostatic SWs (${\bf M} \parallel {\bf k}$) in a magnonic waveguide without the need of an external magnetic field.
Taken together, we hope that this study will motivate SH-wave based magnetoelastic approaches for future applications in magnonics.

\begin{acknowledgments}
This work is financially supported by the German Research Foundation (DFG) via Projects No. WI 1091/21-1, AL 618/36-1, and WE 5386/5-1.
\end{acknowledgments}



\begin{thebibliography}{40}%
\makeatletter
\providecommand \@ifxundefined [1]{%
 \@ifx{#1\undefined}
}%
\providecommand \@ifnum [1]{%
 \ifnum #1\expandafter \@firstoftwo
 \else \expandafter \@secondoftwo
 \fi
}%
\providecommand \@ifx [1]{%
 \ifx #1\expandafter \@firstoftwo
 \else \expandafter \@secondoftwo
 \fi
}%
\providecommand \natexlab [1]{#1}%
\providecommand \enquote  [1]{``#1''}%
\providecommand \bibnamefont  [1]{#1}%
\providecommand \bibfnamefont [1]{#1}%
\providecommand \citenamefont [1]{#1}%
\providecommand \href@noop [0]{\@secondoftwo}%
\providecommand \href [0]{\begingroup \@sanitize@url \@href}%
\providecommand \@href[1]{\@@startlink{#1}\@@href}%
\providecommand \@@href[1]{\endgroup#1\@@endlink}%
\providecommand \@sanitize@url [0]{\catcode `\\12\catcode `\$12\catcode
  `\&12\catcode `\#12\catcode `\^12\catcode `\_12\catcode `\%12\relax}%
\providecommand \@@startlink[1]{}%
\providecommand \@@endlink[0]{}%
\providecommand \url  [0]{\begingroup\@sanitize@url \@url }%
\providecommand \@url [1]{\endgroup\@href {#1}{\urlprefix }}%
\providecommand \urlprefix  [0]{URL }%
\providecommand \Eprint [0]{\href }%
\providecommand \doibase [0]{https://doi.org/}%
\providecommand \selectlanguage [0]{\@gobble}%
\providecommand \bibinfo  [0]{\@secondoftwo}%
\providecommand \bibfield  [0]{\@secondoftwo}%
\providecommand \translation [1]{[#1]}%
\providecommand \BibitemOpen [0]{}%
\providecommand \bibitemStop [0]{}%
\providecommand \bibitemNoStop [0]{.\EOS\space}%
\providecommand \EOS [0]{\spacefactor3000\relax}%
\providecommand \BibitemShut  [1]{\csname bibitem#1\endcsname}%
\let\auto@bib@innerbib\@empty
\bibitem [{\citenamefont {Campbell}(1998)}]{Campbell.1998}%
  \BibitemOpen
  \bibfield  {author} {\bibinfo {author} {\bibfnamefont {C.~K.}\ \bibnamefont
  {Campbell}},\ }\href@noop {} {\emph {\bibinfo {title} {Surface Acoustic Wave
  Devices for Mobile and Wireless Communications}}},\ Applications of Modern
  Acoustics\ (\bibinfo  {publisher} {{Academic Press}},\ \bibinfo {address}
  {San Diego, CA},\ \bibinfo {year} {1998})\BibitemShut {NoStop}%
\bibitem [{\citenamefont {L{\"a}nge}\ \emph {et~al.}(2008)\citenamefont
  {L{\"a}nge}, \citenamefont {Rapp},\ and\ \citenamefont {Rapp}}]{Lange.2008}%
  \BibitemOpen
  \bibfield  {author} {\bibinfo {author} {\bibfnamefont {K.}~\bibnamefont
  {L{\"a}nge}}, \bibinfo {author} {\bibfnamefont {B.~E.}\ \bibnamefont
  {Rapp}},\ and\ \bibinfo {author} {\bibfnamefont {M.}~\bibnamefont {Rapp}},\
  }\bibfield  {title} {\bibinfo {title} {Surface acoustic wave biosensors: a
  review},\ }\href {https://doi.org/10.1007/s00216-008-1911-5} {\bibfield
  {journal} {\bibinfo  {journal} {Anal. Bioanal. Chem.}\ }\textbf {\bibinfo
  {volume} {391}},\ \bibinfo {pages} {1509} (\bibinfo {year}
  {2008})}\BibitemShut {NoStop}%
\bibitem [{\citenamefont {Franke}\ \emph {et~al.}(2009)\citenamefont {Franke},
  \citenamefont {Abate}, \citenamefont {Weitz},\ and\ \citenamefont
  {Wixforth}}]{Franke.2009}%
  \BibitemOpen
  \bibfield  {author} {\bibinfo {author} {\bibfnamefont {T.}~\bibnamefont
  {Franke}}, \bibinfo {author} {\bibfnamefont {A.~R.}\ \bibnamefont {Abate}},
  \bibinfo {author} {\bibfnamefont {D.~A.}\ \bibnamefont {Weitz}},\ and\
  \bibinfo {author} {\bibfnamefont {A.}~\bibnamefont {Wixforth}},\ }\bibfield
  {title} {\bibinfo {title} {Surface acoustic wave ({SAW}) directed droplet
  flow in microfluidics for {PDMS} devices},\ }\href
  {https://doi.org/10.1039/B906819H} {\bibfield  {journal} {\bibinfo  {journal}
  {Lab Chip}\ }\textbf {\bibinfo {volume} {9}},\ \bibinfo {pages} {2625}
  (\bibinfo {year} {2009})}\BibitemShut {NoStop}%
\bibitem [{\citenamefont {{A. Wixforth}}\ \emph {et~al.}(1986)\citenamefont
  {{A. Wixforth}}, \citenamefont {{J. P. Kotthaus}},\ and\ \citenamefont {{G.
  Weimann}}}]{A.Wixforth.1986}%
  \BibitemOpen
  \bibfield  {author} {\bibinfo {author} {\bibnamefont {{A. Wixforth}}},
  \bibinfo {author} {\bibnamefont {{J. P. Kotthaus}}},\ and\ \bibinfo {author}
  {\bibnamefont {{G. Weimann}}},\ }\bibfield  {title} {\bibinfo {title}
  {Quantum oscillations in the surface-acoustic-wave attenuation caused by a
  two-dimensional electron system},\ }\href
  {https://doi.org/10.1103/PhysRevLett.56.2104} {\bibfield  {journal} {\bibinfo
   {journal} {Phys. Rev. Lett.}\ }\textbf {\bibinfo {volume} {56}},\ \bibinfo
  {pages} {2104} (\bibinfo {year} {1986})}\BibitemShut {NoStop}%
\bibitem [{\citenamefont {Fuhrmann}\ \emph {et~al.}(2011)\citenamefont
  {Fuhrmann}, \citenamefont {Thon}, \citenamefont {Kim}, \citenamefont
  {Bouwmeester}, \citenamefont {Petroff}, \citenamefont {Wixforth},\ and\
  \citenamefont {Krenner}}]{DanielA.Fuhrmann.2011}%
  \BibitemOpen
  \bibfield  {author} {\bibinfo {author} {\bibfnamefont {D.~A.}\ \bibnamefont
  {Fuhrmann}}, \bibinfo {author} {\bibfnamefont {S.~M.}\ \bibnamefont {Thon}},
  \bibinfo {author} {\bibfnamefont {H.}~\bibnamefont {Kim}}, \bibinfo {author}
  {\bibfnamefont {D.}~\bibnamefont {Bouwmeester}}, \bibinfo {author}
  {\bibfnamefont {P.~M.}\ \bibnamefont {Petroff}}, \bibinfo {author}
  {\bibfnamefont {A.}~\bibnamefont {Wixforth}},\ and\ \bibinfo {author}
  {\bibfnamefont {H.~J.}\ \bibnamefont {Krenner}},\ }\bibfield  {title}
  {\bibinfo {title} {Dynamic modulation of photonic crystal nanocavities using
  gigahertz acoustic phonons},\ }\href
  {https://doi.org/10.1038/nphoton.2011.208} {\bibfield  {journal} {\bibinfo
  {journal} {Nature Photon}\ }\textbf {\bibinfo {volume} {5}},\ \bibinfo
  {pages} {605} (\bibinfo {year} {2011})}\BibitemShut {NoStop}%
\bibitem [{\citenamefont {{A. Kittmann}}\ \emph {et~al.}(2018)\citenamefont
  {{A. Kittmann}}, \citenamefont {{P. Durdaut}}, \citenamefont {{S. Zabel}},
  \citenamefont {{J. Reermann}}, \citenamefont {{J. Schmalz}}, \citenamefont
  {{Be. Spetzler}}, \citenamefont {{D. Meyners}}, \citenamefont {{N. X. Sun}},
  \citenamefont {{J. McCord}}, \citenamefont {{M. Gerken}}, \citenamefont {{G.
  Schmidt}}, \citenamefont {{M. H{\"o}ft}}, \citenamefont {{R. Kn{\"o}chel}},
  \citenamefont {{F. Faupel}},\ and\ \citenamefont {{E.
  Quandt}}}]{A.Kittmann.2018}%
  \BibitemOpen
  \bibfield  {author} {\bibinfo {author} {\bibnamefont {{A. Kittmann}}},
  \bibinfo {author} {\bibnamefont {{P. Durdaut}}}, \bibinfo {author}
  {\bibnamefont {{S. Zabel}}}, \bibinfo {author} {\bibnamefont {{J.
  Reermann}}}, \bibinfo {author} {\bibnamefont {{J. Schmalz}}}, \bibinfo
  {author} {\bibnamefont {{Be. Spetzler}}}, \bibinfo {author} {\bibnamefont
  {{D. Meyners}}}, \bibinfo {author} {\bibnamefont {{N. X. Sun}}}, \bibinfo
  {author} {\bibnamefont {{J. McCord}}}, \bibinfo {author} {\bibnamefont {{M.
  Gerken}}}, \bibinfo {author} {\bibnamefont {{G. Schmidt}}}, \bibinfo {author}
  {\bibnamefont {{M. H{\"o}ft}}}, \bibinfo {author} {\bibnamefont {{R.
  Kn{\"o}chel}}}, \bibinfo {author} {\bibnamefont {{F. Faupel}}},\ and\
  \bibinfo {author} {\bibnamefont {{E. Quandt}}},\ }\bibfield  {title}
  {\bibinfo {title} {Wide band low noise {L}ove wave magnetic field sensor
  system},\ }\href {https://doi.org/10.1038/s41598-017-18441-4} {\bibfield
  {journal} {\bibinfo  {journal} {Sci. Rep.}\ }\textbf {\bibinfo {volume}
  {8}},\ \bibinfo {pages} {1} (\bibinfo {year} {2018})}\BibitemShut {NoStop}%
\bibitem [{\citenamefont {Weiler}\ \emph {et~al.}(2011)\citenamefont {Weiler},
  \citenamefont {Dreher}, \citenamefont {Heeg}, \citenamefont {Huebl},
  \citenamefont {Gross}, \citenamefont {Brandt},\ and\ \citenamefont
  {Goennenwein}}]{Weiler.2011}%
  \BibitemOpen
  \bibfield  {author} {\bibinfo {author} {\bibfnamefont {M.}~\bibnamefont
  {Weiler}}, \bibinfo {author} {\bibfnamefont {L.}~\bibnamefont {Dreher}},
  \bibinfo {author} {\bibfnamefont {C.}~\bibnamefont {Heeg}}, \bibinfo {author}
  {\bibfnamefont {H.}~\bibnamefont {Huebl}}, \bibinfo {author} {\bibfnamefont
  {R.}~\bibnamefont {Gross}}, \bibinfo {author} {\bibfnamefont {M.~S.}\
  \bibnamefont {Brandt}},\ and\ \bibinfo {author} {\bibfnamefont {S.~T.~B.}\
  \bibnamefont {Goennenwein}},\ }\bibfield  {title} {\bibinfo {title}
  {Elastically driven ferromagnetic resonance in nickel thin films},\ }\href
  {https://doi.org/10.1103/PhysRevLett.106.117601} {\bibfield  {journal}
  {\bibinfo  {journal} {Phys. Rev. Lett.}\ }\textbf {\bibinfo {volume} {106}},\
  \bibinfo {pages} {117601} (\bibinfo {year} {2011})}\BibitemShut {NoStop}%
\bibitem [{\citenamefont {Dreher}\ \emph {et~al.}(2012)\citenamefont {Dreher},
  \citenamefont {Weiler}, \citenamefont {Pernpeintner}, \citenamefont {Huebl},
  \citenamefont {Gross}, \citenamefont {Brandt},\ and\ \citenamefont
  {Goennenwein}}]{Dreher.2012}%
  \BibitemOpen
  \bibfield  {author} {\bibinfo {author} {\bibfnamefont {L.}~\bibnamefont
  {Dreher}}, \bibinfo {author} {\bibfnamefont {M.}~\bibnamefont {Weiler}},
  \bibinfo {author} {\bibfnamefont {M.}~\bibnamefont {Pernpeintner}}, \bibinfo
  {author} {\bibfnamefont {H.}~\bibnamefont {Huebl}}, \bibinfo {author}
  {\bibfnamefont {R.}~\bibnamefont {Gross}}, \bibinfo {author} {\bibfnamefont
  {M.~S.}\ \bibnamefont {Brandt}},\ and\ \bibinfo {author} {\bibfnamefont
  {S.~T.~B.}\ \bibnamefont {Goennenwein}},\ }\bibfield  {title} {\bibinfo
  {title} {Surface acoustic wave driven ferromagnetic resonance in nickel thin
  films: Theory and experiment},\ }\href
  {https://doi.org/10.1103/PhysRevB.86.134415} {\bibfield  {journal} {\bibinfo
  {journal} {Phys. Rev. B}\ }\textbf {\bibinfo {volume} {86}},\ \bibinfo
  {pages} {134415} (\bibinfo {year} {2012})}\BibitemShut {NoStop}%
\bibitem [{\citenamefont {Maekawa}\ and\ \citenamefont
  {Tachiki}(1976)}]{Maekawa.1976}%
  \BibitemOpen
  \bibfield  {author} {\bibinfo {author} {\bibfnamefont {S.}~\bibnamefont
  {Maekawa}}\ and\ \bibinfo {author} {\bibfnamefont {M.}~\bibnamefont
  {Tachiki}},\ }\bibfield  {title} {\bibinfo {title} {Surface acoustic
  attenuation due to surface spin wave in ferro- and antiferromagnets},\
  }\href@noop {} {\bibfield  {journal} {\bibinfo  {journal} {AIP Conf. Proc.}\
  }\textbf {\bibinfo {volume} {29}},\ \bibinfo {pages} {542} (\bibinfo {year}
  {1976})}\BibitemShut {NoStop}%
\bibitem [{\citenamefont {Xu}\ \emph {et~al.}(2020)\citenamefont {Xu},
  \citenamefont {Yamamoto}, \citenamefont {Puebla}, \citenamefont {Baumgaertl},
  \citenamefont {Rana}, \citenamefont {Miura}, \citenamefont {Takahashi},
  \citenamefont {Grundler}, \citenamefont {Maekawa},\ and\ \citenamefont
  {Otani}}]{Xu.2020}%
  \BibitemOpen
  \bibfield  {author} {\bibinfo {author} {\bibfnamefont {M.}~\bibnamefont
  {Xu}}, \bibinfo {author} {\bibfnamefont {K.}~\bibnamefont {Yamamoto}},
  \bibinfo {author} {\bibfnamefont {J.}~\bibnamefont {Puebla}}, \bibinfo
  {author} {\bibfnamefont {K.}~\bibnamefont {Baumgaertl}}, \bibinfo {author}
  {\bibfnamefont {B.}~\bibnamefont {Rana}}, \bibinfo {author} {\bibfnamefont
  {K.}~\bibnamefont {Miura}}, \bibinfo {author} {\bibfnamefont
  {H.}~\bibnamefont {Takahashi}}, \bibinfo {author} {\bibfnamefont
  {D.}~\bibnamefont {Grundler}}, \bibinfo {author} {\bibfnamefont
  {S.}~\bibnamefont {Maekawa}},\ and\ \bibinfo {author} {\bibfnamefont
  {Y.}~\bibnamefont {Otani}},\ }\bibfield  {title} {\bibinfo {title}
  {Nonreciprocal surface acoustic wave propagation via magneto-rotation
  coupling},\ }\href {https://doi.org/10.1126/sciadv.abb1724} {\bibfield
  {journal} {\bibinfo  {journal} {Sci. Adv.}\ }\textbf {\bibinfo {volume}
  {6}},\ \bibinfo {pages} {eabb1724} (\bibinfo {year} {2020})}\BibitemShut
  {NoStop}%
\bibitem [{\citenamefont {Kobayashi}\ \emph {et~al.}(2017)\citenamefont
  {Kobayashi}, \citenamefont {Yoshikawa}, \citenamefont {Matsuo}, \citenamefont
  {Iguchi}, \citenamefont {Maekawa}, \citenamefont {Saitoh},\ and\
  \citenamefont {Nozaki}}]{Kobayashi.2017}%
  \BibitemOpen
  \bibfield  {author} {\bibinfo {author} {\bibfnamefont {D.}~\bibnamefont
  {Kobayashi}}, \bibinfo {author} {\bibfnamefont {T.}~\bibnamefont
  {Yoshikawa}}, \bibinfo {author} {\bibfnamefont {M.}~\bibnamefont {Matsuo}},
  \bibinfo {author} {\bibfnamefont {R.}~\bibnamefont {Iguchi}}, \bibinfo
  {author} {\bibfnamefont {S.}~\bibnamefont {Maekawa}}, \bibinfo {author}
  {\bibfnamefont {E.}~\bibnamefont {Saitoh}},\ and\ \bibinfo {author}
  {\bibfnamefont {Y.}~\bibnamefont {Nozaki}},\ }\bibfield  {title} {\bibinfo
  {title} {Spin current generation using a surface acoustic wave generated via
  spin-rotation coupling},\ }\href
  {https://doi.org/10.1103/PhysRevLett.119.077202} {\bibfield  {journal}
  {\bibinfo  {journal} {Phys. Rev. Lett.}\ }\textbf {\bibinfo {volume} {119}},\
  \bibinfo {pages} {077202} (\bibinfo {year} {2017})}\BibitemShut {NoStop}%
\bibitem [{\citenamefont {Kurimune}\ \emph {et~al.}(2020)\citenamefont
  {Kurimune}, \citenamefont {Matsuo},\ and\ \citenamefont
  {Nozaki}}]{Kurimune.2020}%
  \BibitemOpen
  \bibfield  {author} {\bibinfo {author} {\bibfnamefont {Y.}~\bibnamefont
  {Kurimune}}, \bibinfo {author} {\bibfnamefont {M.}~\bibnamefont {Matsuo}},\
  and\ \bibinfo {author} {\bibfnamefont {Y.}~\bibnamefont {Nozaki}},\
  }\bibfield  {title} {\bibinfo {title} {Observation of gyromagnetic spin wave
  resonance in {NiFe} films},\ }\href
  {https://doi.org/10.1103/PhysRevLett.124.217205} {\bibfield  {journal}
  {\bibinfo  {journal} {Phys. Rev. Lett.}\ }\textbf {\bibinfo {volume} {124}},\
  \bibinfo {pages} {217205} (\bibinfo {year} {2020})}\BibitemShut {NoStop}%
\bibitem [{\citenamefont {Gowtham}\ \emph {et~al.}(2015)\citenamefont
  {Gowtham}, \citenamefont {Moriyama}, \citenamefont {Ralph},\ and\
  \citenamefont {Buhrman}}]{Gowtham.2015}%
  \BibitemOpen
  \bibfield  {author} {\bibinfo {author} {\bibfnamefont {P.~G.}\ \bibnamefont
  {Gowtham}}, \bibinfo {author} {\bibfnamefont {T.}~\bibnamefont {Moriyama}},
  \bibinfo {author} {\bibfnamefont {D.~C.}\ \bibnamefont {Ralph}},\ and\
  \bibinfo {author} {\bibfnamefont {R.~A.}\ \bibnamefont {Buhrman}},\
  }\bibfield  {title} {\bibinfo {title} {Traveling surface spin-wave resonance
  spectroscopy using surface acoustic waves},\ }\href
  {https://doi.org/10.1063/1.4938390} {\bibfield  {journal} {\bibinfo
  {journal} {J. Appl. Phys.}\ }\textbf {\bibinfo {volume} {118}},\ \bibinfo
  {pages} {233910} (\bibinfo {year} {2015})}\BibitemShut {NoStop}%
\bibitem [{\citenamefont {{M. K{\"u}{\ss}}}\ \emph {et~al.}(2020)\citenamefont
  {{M. K{\"u}{\ss}}}, \citenamefont {{M. Heigl}}, \citenamefont {{L. Flacke}},
  \citenamefont {{A. H{\"o}rner}}, \citenamefont {{M. Weiler}}, \citenamefont
  {{M. Albrecht}},\ and\ \citenamefont {{A. Wixforth}}}]{M.Ku.2020}%
  \BibitemOpen
  \bibfield  {author} {\bibinfo {author} {\bibnamefont {{M. K{\"u}{\ss}}}},
  \bibinfo {author} {\bibnamefont {{M. Heigl}}}, \bibinfo {author}
  {\bibnamefont {{L. Flacke}}}, \bibinfo {author} {\bibnamefont {{A.
  H{\"o}rner}}}, \bibinfo {author} {\bibnamefont {{M. Weiler}}}, \bibinfo
  {author} {\bibnamefont {{M. Albrecht}}},\ and\ \bibinfo {author}
  {\bibnamefont {{A. Wixforth}}},\ }\bibfield  {title} {\bibinfo {title}
  {Nonreciprocal {D}zyaloshinskii--{M}oriya magnetoacoustic waves},\ }\href
  {https://doi.org/10.1103/PhysRevLett.125.217203} {\bibfield  {journal}
  {\bibinfo  {journal} {Phys. Rev. Lett.}\ }\textbf {\bibinfo {volume} {125}},\
  \bibinfo {pages} {217203} (\bibinfo {year} {2020})}\BibitemShut {NoStop}%
\bibitem [{\citenamefont {Sasaki}\ \emph {et~al.}(2017)\citenamefont {Sasaki},
  \citenamefont {Nii}, \citenamefont {Iguchi},\ and\ \citenamefont
  {Onose}}]{Sasaki.2017}%
  \BibitemOpen
  \bibfield  {author} {\bibinfo {author} {\bibfnamefont {R.}~\bibnamefont
  {Sasaki}}, \bibinfo {author} {\bibfnamefont {Y.}~\bibnamefont {Nii}},
  \bibinfo {author} {\bibfnamefont {Y.}~\bibnamefont {Iguchi}},\ and\ \bibinfo
  {author} {\bibfnamefont {Y.}~\bibnamefont {Onose}},\ }\bibfield  {title}
  {\bibinfo {title} {Nonreciprocal propagation of surface acoustic wave in
  {Ni}/{LiNbO}$_3$},\ }\href {https://doi.org/10.1103/PhysRevB.95.020407}
  {\bibfield  {journal} {\bibinfo  {journal} {Phys. Rev. B}\ }\textbf {\bibinfo
  {volume} {95}},\ \bibinfo {pages} {020407(R)} (\bibinfo {year}
  {2017})}\BibitemShut {NoStop}%
\bibitem [{\citenamefont {{A. Hern{\'a}ndez-M{\'i}nguez}}\ \emph
  {et~al.}(2020)\citenamefont {{A. Hern{\'a}ndez-M{\'i}nguez}}, \citenamefont
  {{F. Maci{\`a}}}, \citenamefont {{J. M. Hern{\`a}ndez}}, \citenamefont {{J.
  Herfort}},\ and\ \citenamefont {{P. V. Santos}}}]{A.HernandezMinguez.2020}%
  \BibitemOpen
  \bibfield  {author} {\bibinfo {author} {\bibnamefont {{A.
  Hern{\'a}ndez-M{\'i}nguez}}}, \bibinfo {author} {\bibnamefont {{F.
  Maci{\`a}}}}, \bibinfo {author} {\bibnamefont {{J. M. Hern{\`a}ndez}}},
  \bibinfo {author} {\bibnamefont {{J. Herfort}}},\ and\ \bibinfo {author}
  {\bibnamefont {{P. V. Santos}}},\ }\bibfield  {title} {\bibinfo {title}
  {Large nonreciprocal propagation of surface acoustic waves in epitaxial
  ferromagnetic/semiconductor hybrid structures},\ }\href
  {https://doi.org/10.1103/PhysRevApplied.13.044018} {\bibfield  {journal}
  {\bibinfo  {journal} {Phys. Rev. Applied}\ }\textbf {\bibinfo {volume}
  {13}},\ \bibinfo {pages} {044018} (\bibinfo {year} {2020})}\BibitemShut
  {NoStop}%
\bibitem [{\citenamefont {Tateno}\ and\ \citenamefont
  {Nozaki}(2020)}]{Tateno.2020}%
  \BibitemOpen
  \bibfield  {author} {\bibinfo {author} {\bibfnamefont {S.}~\bibnamefont
  {Tateno}}\ and\ \bibinfo {author} {\bibfnamefont {Y.}~\bibnamefont
  {Nozaki}},\ }\bibfield  {title} {\bibinfo {title} {Highly nonreciprocal spin
  waves excited by magnetoelastic coupling in a {Ni}/{Si} bilayer},\ }\href
  {https://doi.org/10.1103/PhysRevApplied.13.034074} {\bibfield  {journal}
  {\bibinfo  {journal} {Phys. Rev. Applied}\ }\textbf {\bibinfo {volume}
  {13}},\ \bibinfo {pages} {034074} (\bibinfo {year} {2020})}\BibitemShut
  {NoStop}%
\bibitem [{\citenamefont {Verba}\ \emph {et~al.}(2018)\citenamefont {Verba},
  \citenamefont {Lisenkov}, \citenamefont {Krivorotov}, \citenamefont
  {Tiberkevich},\ and\ \citenamefont {Slavin}}]{Verba.2018}%
  \BibitemOpen
  \bibfield  {author} {\bibinfo {author} {\bibfnamefont {R.}~\bibnamefont
  {Verba}}, \bibinfo {author} {\bibfnamefont {I.}~\bibnamefont {Lisenkov}},
  \bibinfo {author} {\bibfnamefont {I.}~\bibnamefont {Krivorotov}}, \bibinfo
  {author} {\bibfnamefont {V.}~\bibnamefont {Tiberkevich}},\ and\ \bibinfo
  {author} {\bibfnamefont {A.}~\bibnamefont {Slavin}},\ }\bibfield  {title}
  {\bibinfo {title} {Nonreciprocal surface acoustic waves in multilayers with
  magnetoelastic and interfacial {D}zyaloshinskii-{M}oriya interactions},\
  }\href {https://doi.org/10.1103/PhysRevApplied.9.064014} {\bibfield
  {journal} {\bibinfo  {journal} {Phys. Rev. Applied}\ }\textbf {\bibinfo
  {volume} {9}},\ \bibinfo {pages} {064014} (\bibinfo {year}
  {2018})}\BibitemShut {NoStop}%
\bibitem [{\citenamefont {{R. Verba}}\ \emph {et~al.}(2019)\citenamefont {{R.
  Verba}}, \citenamefont {{V. Tiberkevich}},\ and\ \citenamefont {{A.
  Slavin}}}]{R.Verba.2019}%
  \BibitemOpen
  \bibfield  {author} {\bibinfo {author} {\bibnamefont {{R. Verba}}}, \bibinfo
  {author} {\bibnamefont {{V. Tiberkevich}}},\ and\ \bibinfo {author}
  {\bibnamefont {{A. Slavin}}},\ }\bibfield  {title} {\bibinfo {title}
  {Wide-band nonreciprocity of surface acoustic waves induced by magnetoelastic
  coupling with a synthetic antiferromagnet},\ }\href
  {https://doi.org/10.1103/PhysRevApplied.12.054061} {\bibfield  {journal}
  {\bibinfo  {journal} {Phys. Rev. Applied}\ }\textbf {\bibinfo {volume}
  {12}},\ \bibinfo {pages} {054061} (\bibinfo {year} {2019})}\BibitemShut
  {NoStop}%
\bibitem [{\citenamefont {Labanowski}\ \emph {et~al.}(2016)\citenamefont
  {Labanowski}, \citenamefont {Jung},\ and\ \citenamefont
  {Salahuddin}}]{Labanowski.2016}%
  \BibitemOpen
  \bibfield  {author} {\bibinfo {author} {\bibfnamefont {D.}~\bibnamefont
  {Labanowski}}, \bibinfo {author} {\bibfnamefont {A.}~\bibnamefont {Jung}},\
  and\ \bibinfo {author} {\bibfnamefont {S.}~\bibnamefont {Salahuddin}},\
  }\bibfield  {title} {\bibinfo {title} {Power absorption in acoustically
  driven ferromagnetic resonance},\ }\href {https://doi.org/10.1063/1.4939914}
  {\bibfield  {journal} {\bibinfo  {journal} {Appl. Phys. Lett.}\ }\textbf
  {\bibinfo {volume} {108}},\ \bibinfo {pages} {022905} (\bibinfo {year}
  {2016})}\BibitemShut {NoStop}%
\bibitem [{\citenamefont {Zhou}\ \emph {et~al.}(2014)\citenamefont {Zhou},
  \citenamefont {Talbi}, \citenamefont {Tiercelin},\ and\ \citenamefont
  {Bou~Matar}}]{Zhou.2014}%
  \BibitemOpen
  \bibfield  {author} {\bibinfo {author} {\bibfnamefont {H.}~\bibnamefont
  {Zhou}}, \bibinfo {author} {\bibfnamefont {A.}~\bibnamefont {Talbi}},
  \bibinfo {author} {\bibfnamefont {N.}~\bibnamefont {Tiercelin}},\ and\
  \bibinfo {author} {\bibfnamefont {O.}~\bibnamefont {Bou~Matar}},\ }\bibfield
  {title} {\bibinfo {title} {Multilayer magnetostrictive structure based
  surface acoustic wave devices},\ }\href {https://doi.org/10.1063/1.4868530}
  {\bibfield  {journal} {\bibinfo  {journal} {Appl. Phys. Lett.}\ }\textbf
  {\bibinfo {volume} {104}},\ \bibinfo {pages} {114101} (\bibinfo {year}
  {2014})}\BibitemShut {NoStop}%
\bibitem [{\citenamefont {{A. Mazzamurro}}\ \emph {et~al.}(2020)\citenamefont
  {{A. Mazzamurro}}, \citenamefont {{Y. Dusch}}, \citenamefont {{P. Pernod}},
  \citenamefont {{O. Bou Matar}}, \citenamefont {{A. Addad}}, \citenamefont
  {{A. Talbi}},\ and\ \citenamefont {{N. Tiercelin}}}]{A.Mazzamurro.2020}%
  \BibitemOpen
  \bibfield  {author} {\bibinfo {author} {\bibnamefont {{A. Mazzamurro}}},
  \bibinfo {author} {\bibnamefont {{Y. Dusch}}}, \bibinfo {author}
  {\bibnamefont {{P. Pernod}}}, \bibinfo {author} {\bibnamefont {{O. Bou
  Matar}}}, \bibinfo {author} {\bibnamefont {{A. Addad}}}, \bibinfo {author}
  {\bibnamefont {{A. Talbi}}},\ and\ \bibinfo {author} {\bibnamefont {{N.
  Tiercelin}}},\ }\bibfield  {title} {\bibinfo {title} {Giant magnetoelastic
  coupling in a {L}ove acoustic waveguide based on {TbCo}$_2$/{FeCo}
  nanostructured film on {ST}-cut quartz},\ }\href
  {https://doi.org/10.1103/PhysRevApplied.13.044001} {\bibfield  {journal}
  {\bibinfo  {journal} {Phys. Rev. Applied}\ }\textbf {\bibinfo {volume}
  {13}},\ \bibinfo {pages} {044001} (\bibinfo {year} {2020})}\BibitemShut
  {NoStop}%
\bibitem [{\citenamefont {Morgan}(2007)}]{Morgan.2007}%
  \BibitemOpen
  \bibfield  {author} {\bibinfo {author} {\bibfnamefont {D.~P.}\ \bibnamefont
  {Morgan}},\ }\href
  {http://site.ebrary.com/lib/alltitles/docDetail.action?docID=10186713} {\emph
  {\bibinfo {title} {Surface Acoustic Wave Filters: With Applications to
  Electronic Communications and Signal Processing}}},\ \bibinfo {edition}
  {2nd}\ ed.\ (\bibinfo  {publisher} {Elsevier},\ \bibinfo {address}
  {Amsterdam},\ \bibinfo {year} {2007})\BibitemShut {NoStop}%
\bibitem [{\citenamefont {Nakamura}\ \emph {et~al.}(1977)\citenamefont
  {Nakamura}, \citenamefont {Kazumi},\ and\ \citenamefont
  {Shimizu}}]{Nakamura.1977}%
  \BibitemOpen
  \bibfield  {author} {\bibinfo {author} {\bibfnamefont {K.}~\bibnamefont
  {Nakamura}}, \bibinfo {author} {\bibfnamefont {M.}~\bibnamefont {Kazumi}},\
  and\ \bibinfo {author} {\bibfnamefont {H.}~\bibnamefont {Shimizu}},\
  }\bibfield  {title} {\bibinfo {title} {{SH}-type and {R}ayleigh-type surface
  waves on rotated {Y}-cut {LiTaO}$_3$}\ }(\bibinfo  {publisher} {IEEE},\
  \bibinfo {address} {Phoenix},\ \bibinfo {year} {1977})\ pp.\ \bibinfo {pages}
  {819--822}\BibitemShut {NoStop}%
\bibitem [{\citenamefont {Matsuo}\ \emph {et~al.}(2011)\citenamefont {Matsuo},
  \citenamefont {Ieda}, \citenamefont {Saitoh},\ and\ \citenamefont
  {Maekawa}}]{Matsuo.2011}%
  \BibitemOpen
  \bibfield  {author} {\bibinfo {author} {\bibfnamefont {M.}~\bibnamefont
  {Matsuo}}, \bibinfo {author} {\bibfnamefont {J.}~\bibnamefont {Ieda}},
  \bibinfo {author} {\bibfnamefont {E.}~\bibnamefont {Saitoh}},\ and\ \bibinfo
  {author} {\bibfnamefont {S.}~\bibnamefont {Maekawa}},\ }\bibfield  {title}
  {\bibinfo {title} {Effects of mechanical rotation on spin currents},\ }\href
  {https://doi.org/10.1103/PhysRevLett.106.076601} {\bibfield  {journal}
  {\bibinfo  {journal} {Phys. Rev. Lett.}\ }\textbf {\bibinfo {volume} {106}},\
  \bibinfo {pages} {076601} (\bibinfo {year} {2011})}\BibitemShut {NoStop}%
\bibitem [{\citenamefont {Matsuo}\ \emph {et~al.}(2013)\citenamefont {Matsuo},
  \citenamefont {Ieda}, \citenamefont {Harii}, \citenamefont {Saitoh},\ and\
  \citenamefont {Maekawa}}]{Matsuo.2013}%
  \BibitemOpen
  \bibfield  {author} {\bibinfo {author} {\bibfnamefont {M.}~\bibnamefont
  {Matsuo}}, \bibinfo {author} {\bibfnamefont {J.}~\bibnamefont {Ieda}},
  \bibinfo {author} {\bibfnamefont {K.}~\bibnamefont {Harii}}, \bibinfo
  {author} {\bibfnamefont {E.}~\bibnamefont {Saitoh}},\ and\ \bibinfo {author}
  {\bibfnamefont {S.}~\bibnamefont {Maekawa}},\ }\bibfield  {title} {\bibinfo
  {title} {Mechanical generation of spin current by spin-rotation coupling},\
  }\href {https://doi.org/10.1103/PhysRevB.87.180402} {\bibfield  {journal}
  {\bibinfo  {journal} {Phys. Rev. B}\ }\textbf {\bibinfo {volume} {87}},\
  \bibinfo {pages} {180402(R)} (\bibinfo {year} {2013})}\BibitemShut {NoStop}%
\bibitem [{\citenamefont {Kalinikos}\ and\ \citenamefont
  {Slavin}(1986)}]{Kalinikos.}%
  \BibitemOpen
  \bibfield  {author} {\bibinfo {author} {\bibfnamefont {B.~A.}\ \bibnamefont
  {Kalinikos}}\ and\ \bibinfo {author} {\bibfnamefont {A.~N.}\ \bibnamefont
  {Slavin}},\ }\bibfield  {title} {\bibinfo {title} {Theory of dipole-exchange
  spin wave spectrum for ferromagnetic films with mixed exchange boundary
  conditions},\ }\href {https://doi.org/10.1088/0022-3719/19/35/014} {\bibfield
   {journal} {\bibinfo  {journal} {J. Phys. C}\ }\textbf {\bibinfo {volume}
  {19}},\ \bibinfo {pages} {7013} (\bibinfo {year} {1986})}\BibitemShut
  {NoStop}%
\bibitem [{Com()}]{Comsol.}%
  \BibitemOpen
  \href@noop {} {}\bibinfo {note} {COMSOL Multiphysics{$^\text{\circledR}$} v.
  5.4. \url{www.comsol.com.} COMSOL AB, Stockholm, Sweden.}\BibitemShut {Stop}%
\bibitem [{\citenamefont {Mills}(1999)}]{Mills.1999}%
  \BibitemOpen
  \bibfield  {author} {\bibinfo {author} {\bibfnamefont {A.~F.}\ \bibnamefont
  {Mills}},\ }\href {https://cds.cern.ch/record/2291687} {\emph {\bibinfo
  {title} {Basic Heat and Mass Transfer}}}\ (\bibinfo  {publisher} {{Prentice
  Hall}},\ \bibinfo {address} {Upper Saddle River, NJ},\ \bibinfo {year}
  {1999})\BibitemShut {NoStop}%
\bibitem [{\citenamefont {Comte}\ and\ \citenamefont {von
  Stebut}(2002)}]{Comte.2002}%
  \BibitemOpen
  \bibfield  {author} {\bibinfo {author} {\bibfnamefont {C.}~\bibnamefont
  {Comte}}\ and\ \bibinfo {author} {\bibfnamefont {J.}~\bibnamefont {von
  Stebut}},\ }\bibfield  {title} {\bibinfo {title} {Microprobe-type measurement
  of {Y}oung's modulus and {P}oisson coefficient by means of depth sensing
  indentation and acoustic microscopy},\ }\href
  {https://doi.org/10.1016/S0257-8972(01)01706-6} {\bibfield  {journal}
  {\bibinfo  {journal} {Surf. Coat. Technol.}\ }\textbf {\bibinfo {volume}
  {154}},\ \bibinfo {pages} {42} (\bibinfo {year} {2002})}\BibitemShut
  {NoStop}%
\bibitem [{\citenamefont {Ledbetter}\ and\ \citenamefont
  {Reed}(1973)}]{Ledbetter.1973}%
  \BibitemOpen
  \bibfield  {author} {\bibinfo {author} {\bibfnamefont {H.~M.}\ \bibnamefont
  {Ledbetter}}\ and\ \bibinfo {author} {\bibfnamefont {R.~P.}\ \bibnamefont
  {Reed}},\ }\bibfield  {title} {\bibinfo {title} {Elastic properties of metals
  and alloys, {I}. iron, nickel, and iron-nickel alloys},\ }\href
  {https://doi.org/10.1063/1.3253127} {\bibfield  {journal} {\bibinfo
  {journal} {J. Phys. Chem. Ref. Data}\ }\textbf {\bibinfo {volume} {2}},\
  \bibinfo {pages} {531} (\bibinfo {year} {1973})}\BibitemShut {NoStop}%
\bibitem [{\citenamefont {Ekstr{\"o}m}\ \emph {et~al.}(2017)\citenamefont
  {Ekstr{\"o}m}, \citenamefont {Aref}, \citenamefont {Runeson}, \citenamefont
  {Bj{\"o}rck}, \citenamefont {Bostr{\"o}m},\ and\ \citenamefont
  {Delsing}}]{Ekstrom.2017}%
  \BibitemOpen
  \bibfield  {author} {\bibinfo {author} {\bibfnamefont {M.~K.}\ \bibnamefont
  {Ekstr{\"o}m}}, \bibinfo {author} {\bibfnamefont {T.}~\bibnamefont {Aref}},
  \bibinfo {author} {\bibfnamefont {J.}~\bibnamefont {Runeson}}, \bibinfo
  {author} {\bibfnamefont {J.}~\bibnamefont {Bj{\"o}rck}}, \bibinfo {author}
  {\bibfnamefont {I.}~\bibnamefont {Bostr{\"o}m}},\ and\ \bibinfo {author}
  {\bibfnamefont {P.}~\bibnamefont {Delsing}},\ }\bibfield  {title} {\bibinfo
  {title} {Surface acoustic wave unidirectional transducers for quantum
  applications},\ }\href {https://doi.org/10.1063/1.4975803} {\bibfield
  {journal} {\bibinfo  {journal} {Appl. Phys. Lett.}\ }\textbf {\bibinfo
  {volume} {110}},\ \bibinfo {pages} {073105} (\bibinfo {year}
  {2017})}\BibitemShut {NoStop}%
\bibitem [{\citenamefont {Hiebel}(2011)}]{Hiebel.2011}%
  \BibitemOpen
  \bibfield  {author} {\bibinfo {author} {\bibfnamefont {M.}~\bibnamefont
  {Hiebel}},\ }\href@noop {} {\emph {\bibinfo {title} {Grundlagen der
  vektoriellen Netzwerkanalyse}}},\ \bibinfo {edition} {3rd}\ ed.\ (\bibinfo
  {publisher} {{Rohde {\&} Schwarz}},\ \bibinfo {address} {M{\"u}nchen},\
  \bibinfo {year} {2011})\BibitemShut {NoStop}%
\bibitem [{\citenamefont {Wilts}\ and\ \citenamefont {Lai}(1972)}]{Wilts.1972}%
  \BibitemOpen
  \bibfield  {author} {\bibinfo {author} {\bibfnamefont {C.~H.}\ \bibnamefont
  {Wilts}}\ and\ \bibinfo {author} {\bibfnamefont {S.~K.~C.}\ \bibnamefont
  {Lai}},\ }\bibfield  {title} {\bibinfo {title} {Spin wave measurements of
  exchange constant in {N}i-{F}e alloy films},\ }\href
  {https://doi.org/10.1109/TMAG.1972.1067520} {\bibfield  {journal} {\bibinfo
  {journal} {IEEE Trans. Magn.}\ }\textbf {\bibinfo {volume} {8}},\ \bibinfo
  {pages} {280} (\bibinfo {year} {1972})}\BibitemShut {NoStop}%
\bibitem [{Est()}]{EstimateStrain.0001}%
  \BibitemOpen
  \href@noop {} {}\bibinfo {note} {To estimate the approximate amplitude of the
  induced strain $\epsilon_{xx,0}$ for the R-wave ($\epsilon_{xy,0}$ for the
  SH-wave), we assume both IDTs having the same efficiency in SAW excitation
  and detection. Since the delayline has a total insertion loss of about
  $82$~dB for the R-mode ($58$~dB for the SH-mode), the loss to excite the SAW
  in the middle of the delayline is approximately $-41$~dB ($-29$~dB). Because
  the output power of the vector network analyzer is 15~dBm, the power of the
  R-mode (SH-mode) in the middle of the delayline is $P_\text{SAW}
  \approx-26$~dBm ($P_\text{SAW}\approx-14$~dBm). Together with Eq.~(1) of
  Ref.~\cite{Croset.2017} and the normalized displacement profile ${\bf
  u}(x,z)$ from the FEM simulation we estimate the strain $\epsilon_{xx,0}$
  ($\epsilon_{xy,0}$) being in the order of $12 \times 10^{-6}$ ($23 \times
  10^{-6}$).}\BibitemShut {Stop}%
\bibitem [{\citenamefont {Croset}\ \emph {et~al.}(2017)\citenamefont {Croset},
  \citenamefont {Camara}, \citenamefont {Duquesne}, \citenamefont {Largeau},
  \citenamefont {Thevenard},\ and\ \citenamefont {Rovillain}}]{Croset.2017}%
  \BibitemOpen
  \bibfield  {author} {\bibinfo {author} {\bibfnamefont {B.}~\bibnamefont
  {Croset}}, \bibinfo {author} {\bibfnamefont {I.~S.}\ \bibnamefont {Camara}},
  \bibinfo {author} {\bibfnamefont {J.-Y.}\ \bibnamefont {Duquesne}}, \bibinfo
  {author} {\bibfnamefont {L.}~\bibnamefont {Largeau}}, \bibinfo {author}
  {\bibfnamefont {L.}~\bibnamefont {Thevenard}},\ and\ \bibinfo {author}
  {\bibfnamefont {P.}~\bibnamefont {Rovillain}},\ }\bibfield  {title} {\bibinfo
  {title} {Vector network analyzer measurement of the amplitude of an
  electrically excited surface acoustic wave and validation by x-ray
  diffraction},\ }\href {https://doi.org/10.1063/1.4974947} {\bibfield
  {journal} {\bibinfo  {journal} {J. Appl. Phys.}\ }\textbf {\bibinfo {volume}
  {121}},\ \bibinfo {pages} {044503} (\bibinfo {year} {2017})}\BibitemShut
  {NoStop}%
\bibitem [{Oth()}]{OtherRef.0002}%
  \BibitemOpen
  \href@noop {} {}\bibinfo {note} {To estimate the magnetization precession
  amplitude $\phi_\text{M,ip}$, we expanded the theory of
  Ref.~\cite{Dreher.2012} by the dipolar fields of the
  SW~\cite{Kalinikos.,M.Ku.2020}. The parameters which were used for the
  estimation are (i) the strain components as discussed before, (ii) the
  isotropic magnetoelastic coupling constant being $b_1 = b_2 =
  23$~T~\cite{Dreher.2012}, and (iii) the magnetic film properties, given in
  Table~\ref{tab:table2}.}\BibitemShut {Stop}%
\bibitem [{\citenamefont {Casals}\ \emph {et~al.}(2020)\citenamefont {Casals},
  \citenamefont {Statuto}, \citenamefont {Foerster}, \citenamefont
  {Hern{\'a}ndez-M{\'i}nguez}, \citenamefont {Cichelero}, \citenamefont
  {Manshausen}, \citenamefont {Mandziak}, \citenamefont {Aballe}, \citenamefont
  {Hern{\`a}ndez},\ and\ \citenamefont {Maci{\`a}}}]{Casals.2020}%
  \BibitemOpen
  \bibfield  {author} {\bibinfo {author} {\bibfnamefont {B.}~\bibnamefont
  {Casals}}, \bibinfo {author} {\bibfnamefont {N.}~\bibnamefont {Statuto}},
  \bibinfo {author} {\bibfnamefont {M.}~\bibnamefont {Foerster}}, \bibinfo
  {author} {\bibfnamefont {A.}~\bibnamefont {Hern{\'a}ndez-M{\'i}nguez}},
  \bibinfo {author} {\bibfnamefont {R.}~\bibnamefont {Cichelero}}, \bibinfo
  {author} {\bibfnamefont {P.}~\bibnamefont {Manshausen}}, \bibinfo {author}
  {\bibfnamefont {A.}~\bibnamefont {Mandziak}}, \bibinfo {author}
  {\bibfnamefont {L.}~\bibnamefont {Aballe}}, \bibinfo {author} {\bibfnamefont
  {J.~M.}\ \bibnamefont {Hern{\`a}ndez}},\ and\ \bibinfo {author}
  {\bibfnamefont {F.}~\bibnamefont {Maci{\`a}}},\ }\bibfield  {title} {\bibinfo
  {title} {Generation and imaging of magnetoacoustic waves over millimeter
  distances},\ }\href {https://doi.org/10.1103/PhysRevLett.124.137202}
  {\bibfield  {journal} {\bibinfo  {journal} {Phys. Rev. Lett.}\ }\textbf
  {\bibinfo {volume} {124}},\ \bibinfo {pages} {137202} (\bibinfo {year}
  {2020})}\BibitemShut {NoStop}%
\bibitem [{\citenamefont {Azovtsev}\ and\ \citenamefont
  {Pertsev}(2019)}]{Azovtsev.2019}%
  \BibitemOpen
  \bibfield  {author} {\bibinfo {author} {\bibfnamefont {A.~V.}\ \bibnamefont
  {Azovtsev}}\ and\ \bibinfo {author} {\bibfnamefont {N.~A.}\ \bibnamefont
  {Pertsev}},\ }\bibfield  {title} {\bibinfo {title} {Dynamical spin phenomena
  generated by longitudinal elastic waves traversing {CoFe}$_2${O}$_4$ films
  and heterostructures},\ }\href {https://doi.org/10.1103/PhysRevB.100.224405}
  {\bibfield  {journal} {\bibinfo  {journal} {Phys. Rev. B}\ }\textbf {\bibinfo
  {volume} {100}},\ \bibinfo {pages} {224405} (\bibinfo {year}
  {2019})}\BibitemShut {NoStop}%
\bibitem [{\citenamefont {{A. A. Serga}}\ \emph {et~al.}(2010)\citenamefont
  {{A. A. Serga}}, \citenamefont {{A. V. Chumak}},\ and\ \citenamefont {{B.
  Hillebrands}}}]{AASerga.2010}%
  \BibitemOpen
  \bibfield  {author} {\bibinfo {author} {\bibnamefont {{A. A. Serga}}},
  \bibinfo {author} {\bibnamefont {{A. V. Chumak}}},\ and\ \bibinfo {author}
  {\bibnamefont {{B. Hillebrands}}},\ }\bibfield  {title} {\bibinfo {title}
  {{YIG} magnonics},\ }\href {https://doi.org/10.1088/0022-3727/43/26/264002}
  {\bibfield  {journal} {\bibinfo  {journal} {J. Phys. D: Appl. Phys.}\
  }\textbf {\bibinfo {volume} {43}},\ \bibinfo {pages} {264002} (\bibinfo
  {year} {2010})}\BibitemShut {NoStop}%
\end{thebibliography}


%

\end{document}